\documentclass[journal,twoside,final]{IEEEtran}
\usepackage{amsmath}
\usepackage{amsfonts}
\usepackage{sansmath}

\usepackage{algorithm}
\usepackage{algpseudocode}

\usepackage[dvipsnames]{xcolor}

\PassOptionsToPackage{hyphens}{url}
\usepackage[colorlinks]{hyperref}
\hypersetup{
    linkcolor=blue,
    filecolor=magenta,
    urlcolor=teal,
    citecolor=magenta,
    breaklinks = true
    }
\usepackage{breakurl}
\usepackage[noadjust]{cite}

\usepackage{graphicx} % for figures
\usepackage{subfigure} 

\usepackage{bm} % for bold symbols in equations
\usepackage{bbm} % for mathbbm (e.g. the indicator function)

\usepackage{array,multirow} % for the table that uses multirows

\graphicspath{ {./Figs/} }

 \def\Gc{\mathcal G}
 \def\Pc{\mathcal P}
 \def\Nc{\mathcal N}
 \def\Tau{\mathcal T}

 \def\mub{\bm{\mu}}

\def\ssb{\bm{s}}
\def\sigmab{\bm{\varsigma}}

\def\xb{\bm{x}}
\def\xib{\bm{\xi}}

\def\nb{\bm{n}}

\def\etab{\bm{\eta}}
\def\gb{\bm{g}}
\def\Gb{\bm{G}}

\def\Thetab{\bm{\Theta}}
\def\Psib{\bm{\Psi}}

\def\vh{\hat{v}}
 \def\diag#1{\mbox{diag} #1}
\def\eqdef{\stackrel{\Delta}{=}}
\def\Oc{\mathcal O}

\def\zerob{\bm{0}}
\def\oneb{\bm{1}}

\def\Kb{\bm{K}}
\def\kb{\bm{k}}
 \def\ssb{\bm{s}}
 \def\sbh{\widehat{\ssb}}	
\def\diag#1{ \mbox{diag} \left[ #1\right] }
\begin{document}

\title{A Data-Driven Gaussian Process Filter for Electrocardiogram Denoising} % with Applications in QT-interval Estimation

\author{
Mircea~Dumitru,
\and 
Qiao~Li,
\and
Erick~Andres~Perez~Alday,
\and
Ali~Bahrami~Rad,
\and
Gari~D.~Clifford,
\and
Reza~Sameni$^\textnormal{*}$
\thanks{The authors are with the Department of Biomedical Informatics, School of Medicine, Emory University. G. D. Clifford is also with the Biomedical Engineering Department, Georgia Institute of Technology. Corresponding author: R.~Sameni (email: \url{rsameni@dbmi.emory.edu}).
}
}

%////////////////////////////////
%\markboth{46th Conf. of the International Society for Computerized Electrocardiology (ISCE), Apr 6--10, 2022, Las Vegas, US $\quad$[Preprint]}{A Data-Driven ECG Gaussian Process Filter, M. Dumitru et al.}
%\markboth{[Preprint submitted to the] IEEE Transactions on Biomedical Engineering Letters (TBME Letters) $\quad$}{A Data-Driven Gaussian Process ECG Filter, M. Dumitru et al.}
\markboth{Presented at the 46th International Society for Computerized Electrocardiology Conference (ISCE), Apr 6--10, 2022, Las Vegas, US}{A Data-Driven ECG Gaussian Process Filter, M. Dumitru et al.}
%\pubid{0000--0000/00\$00.00~\copyright~2007 IEEE}
\maketitle
%////////////////////////////////
\begin{abstract}
\textit{Objective:} Gaussian Processes ($\Gc\Pc$)-based filters, which have been effectively used for various applications including electrocardiogram (ECG) filtering can be computationally demanding and the choice of their hyperparameters is typically ad hoc. \textit{Methods:} We develop a data-driven $\Gc\Pc$ filter to address both issues, using the notion of the ECG \textit{phase domain} --- a time-warped representation of the ECG beats onto a fixed number of samples and aligned R-peaks, which is assumed to follow a Gaussian distribution. Under this assumption, the computation of the sample mean and covariance matrix is simplified, enabling an efficient implementation of the $\Gc\Pc$ filter in a data-driven manner, with no ad hoc hyperparameters. The proposed filter is evaluated and compared with a state-of-the-art wavelet-based filter, on the PhysioNet QT Database. The performance is evaluated by measuring the signal-to-noise ratio (SNR) improvement of the filter at SNR levels ranging from --5 to 30\,dB, in 5\,dB steps, using additive noise. For a clinical evaluation, the error between the estimated QT-intervals of the original and filtered signals is measured and compared with the benchmark filter. \textit{Results:} It is shown that the proposed $\Gc\Pc$ filter outperforms the benchmark filter for all the tested noise levels. It also outperforms the state-of-the-art filter in terms of QT-interval estimation error bias and variance. \textit{Conclusion:} The proposed $\Gc\Pc$ filter is a versatile technique for preprocessing the ECG in clinical and research applications, is applicable to ECG of arbitrary lengths and sampling frequencies, and provides confidence intervals for its performance.
\end{abstract}

%////////////////////////////////
\begin{IEEEkeywords}
ECG Bayesian filter, Gaussian processes, ECG denoising, ECG wavelet denoising, QT-interval estimation
\end{IEEEkeywords}
%////////////////////////////////
\IEEEpeerreviewmaketitle

%Notes to myself
%2. Tables, caption on top 
%3. ref 9 changes 
%OSET package, at Pant tompking ref
%5. figure 3,4 - font axis 
%

% Section 1 : Introduction
\section{Introduction}
\label{sec:intro}
Electrocardiogram (ECG) denoising is a recurrent problem in traditional and wearable cardiac monitors. The problem has been addressed by various approaches, including model-based and non-model-based filters. A powerful non-parametric framework for ECG filtering is via Gaussian process ($\Gc\Pc$) models \cite{Rivet2012,Niknazar2012}, which considers the ECG beats as $\Gc\Pc$s with common parameters. The choice of the beats $\Gc\Pc$ hyperparameters, namely the mean and kernel functions is non-evident and ad hoc. For $\Gc\Pc$ models with no beat assumptions, beside the ambiguity in parameter selection, the $\Gc\Pc$ filter implementation involves the inversion of large covariance matrices, which precludes the use of this framework for long ECG records.

In this paper, ECG filtering is addressed via a data-driven non-parametric $\Gc\Pc$ model. The novelty of the proposed filter is that it requires no ad hoc $\Gc\Pc$ model hyperparameters and it is computationally efficient, making it suitable for any length ECG records; it is based on the assumption that each \textit{phase domain} beat --- a time-warped (stretched or squeezed)
representation of the ECG beats onto a fixed number of samples
and aligned R-peaks --- is an ensemble of an underlying $\Gc\Pc$. The mean and the kernel function are set via the phase domain sample mean and covariance matrix, computed via the available ensembles, which are transformed back to the time-domain and used to derive the posterior mean using the Bayesian formalism.

This proposed filter is data-driven, does not presume any parametric model for the underlying $\Gc\Pc$, and is computationally efficient.

The filter is evaluated in terms of signal-to-noise ratio (SNR) improvement, using as benchmark a wavelet-based ECG denoiser that was demonstrated in \cite{Sameni2017} to outperform adaptive filters \cite{Laguna1992}, Tikhonov regularization and Extended Kalman filters \cite{SSJC06}, in terms of SNR improvement. The proposed filter's clinical performance is evaluated by measuring the QT-interval error between the clean ECG and its corresponding filtered version. 

% Section 2 : GP based model
\section{Gaussian process-based ECG filtering}
\label{sec:GP-model}
\subsection{The mathematical model}
\label{subsec:the-math-model}
The ECG measurement $x(t)$ is assumed to be an additive mixture of a clean ECG $s(t)$, assumed to be a $\Gc\Pc$ contaminated by additive white noise:
\begin{equation}
\label{eeq:1}
    x(t)
    =
    s(t)
    +
    n(t)
    ,
    \quad
    t \in \left\lbrace t_1 \ldots t_N \right\rbrace \eqdef T_{N},
\end{equation}
where $n(t) \sim\mathcal{N}(0, v_n)$, $v_n$ denotes the noise variance and $N$ denotes the number of measurements. The signal $x(t)$ is assumed to be baseline-wander (BW) and powerline noise removed, which are relatively straightforward, with classical filtering pipelines (cf. Section \ref{subsec:BW-removal}). Therefore, the filter design objective is focused on \textit{in-band ECG noise} removal.

% Finally it is also assumed that the clean ECG is a $\Gc\Pc$ and the noise is a zero-mean white $\Gc\Pc$ with variance $v_{n}$:
% \begin{equation}
% \label{eeq:2}
% \begin{split}
%     s(t)
%     %&
%     \sim
%     \GP
%     {\mu_s(t)}
%     {\ker{\srv}{t}{t'}}
%     ,
%     %\\
%     n(t)
%     %&
%     \sim
%     \GP
%     {0}
%     {v_{n} \delta\left(t - t' \right)}    
% \end{split}
% \end{equation}
% where $\delta(\cdot)$ is the Dirac delta function. 

For the beat $i$, $T_{i} = \left\lbrace t_{i_1} \ldots t_{i_{R_i}} \ldots t_{i_{N_i}} \right\rbrace$ denotes the set of time samples, $t_{i_1}$ representing the first sample, $t_{i_{R_i}}$ the sample corresponding to the R-peak and $t_{i_{N_i}}$ the last sample. We further define $\xb_i = [x(t)]_{i\in T_i}$, $\ssb_i = [s(t)]_{i\in T_i}$, $\nb_i = [n(t)]_{i\in T_i}$ as vectorial representations of the measurement, clean ECG and noise, respectively. Therefore, $\xb_i = \ssb_i + \nb_i$.
% from \eqref{eeq:2}
% \begin{equation}
% \label{eeq:3}
%   \xb_i = \ssb_i + \nb_i
%   ,
%   \;\;
%   \nb_i \sim \Nc(\zerob, v_n \Ib_{N_i})
%   ,
%   \;\;
%   \ssb_i \sim \Nc(\mub_{s_i}, \Kb_{s_i})
% \end{equation}

\begin{figure}
\centering
\includegraphics[width=0.95\columnwidth]{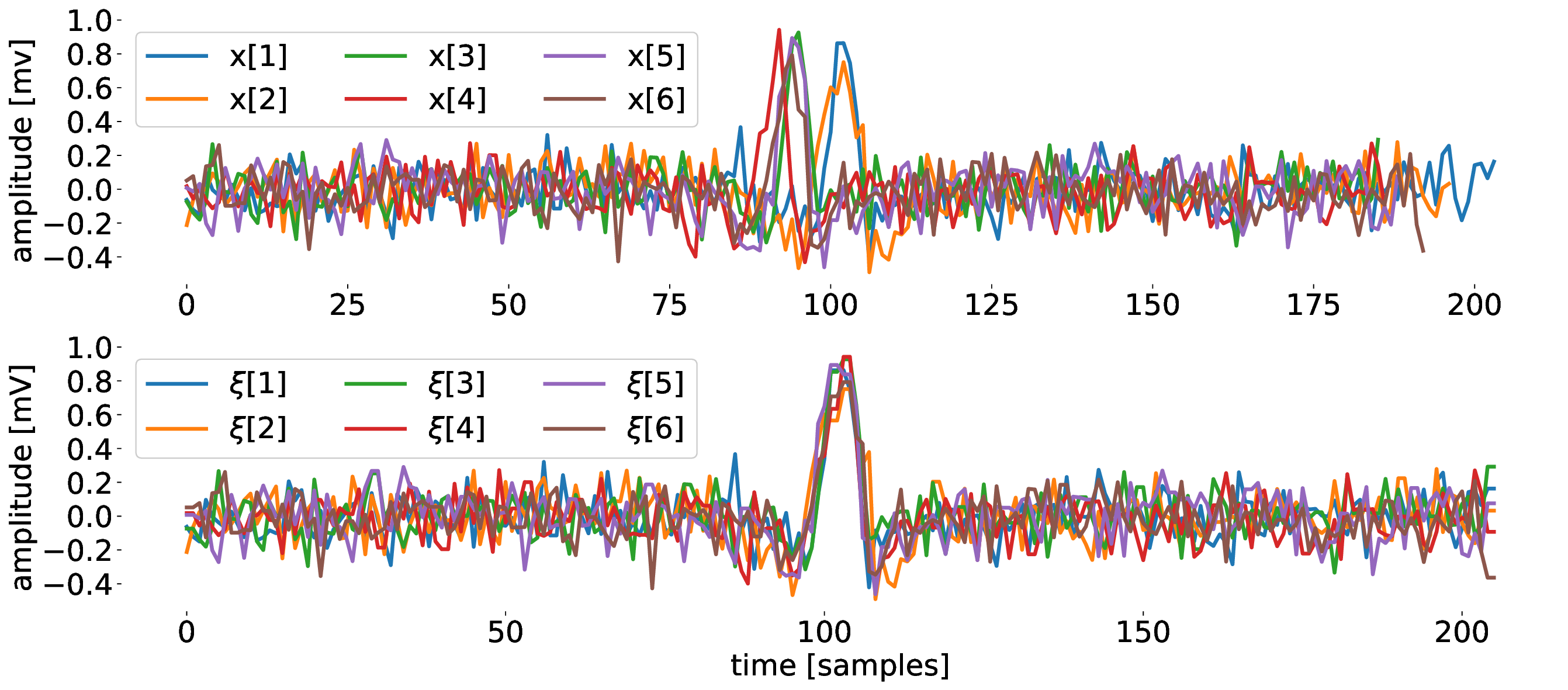}
\caption{Time-domain measurements beats (top) and the corresponding phase domain ECG beats (bottom), with the same number $\Tau$ of samples for the first 6 beats of \texttt{sel100} record from QTDB \cite{Laguna1997} with $0$\,dB Gaussian additive noise. Transformation matrices $\Thetab_{i}$ are defined via \eqref{eeq:2bis}.}
\label{fig:time-and-beats}
\end{figure}

\begin{figure}
\centering
\includegraphics[width=0.95\columnwidth]{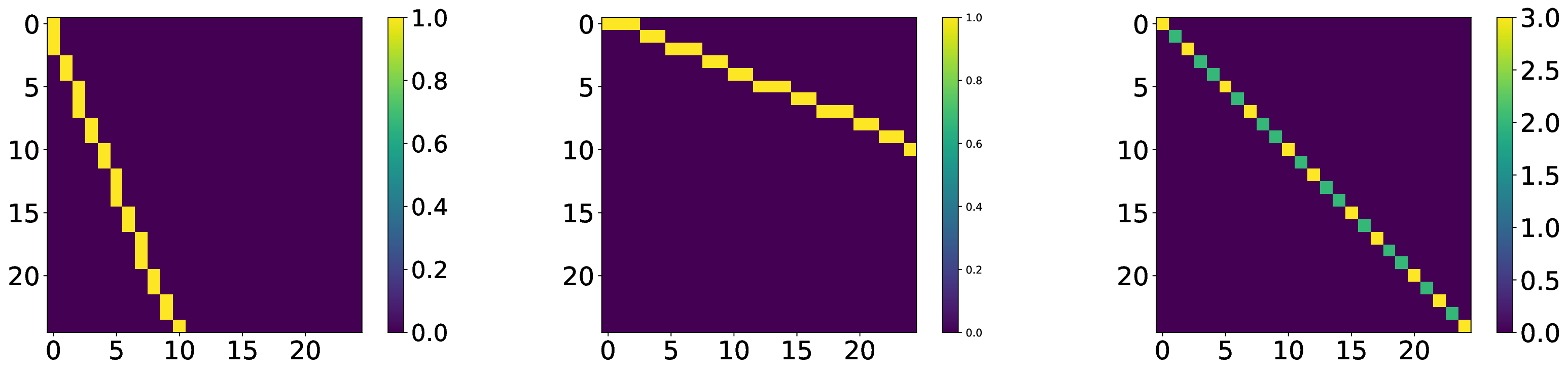}
\caption{Corner detail example of transformation matrix $\Thetab_{i}$ (left), $\Thetab_{i}^{T}$ (middle) and the corresponding (diagonal) Gramian $\Gb_{i} = \Thetab_{i}^{T} \Thetab_{i}$ (right).}
\label{fig:transf-matrix}
\end{figure}

Next, we define matrices $\Thetab_i \in \mathbb{R}^{\Tau \times N_i}$ to map the time domain beats $\xb_i$, $\ssb_i$ and $\nb_i$ to the phase domain beats 
\begin{equation}
\label{eeq:2}
\xib_i = \Thetab_i \xb_i
, 
\;
\;
\sigmab_i = \Thetab_i \ssb_i 
,
\;
\;
\etab_i = \Thetab_i \nb_i
,
\end{equation}
with aligned R-peaks and the same number of samples $\Tau$ (Fig.~\ref{fig:time-and-beats}).
The $\Thetab_i$ matrices are defined by considering $\Tau$ knots equidistantly distributed in the interval $[1,N_i]$ and assigning 
\begin{equation}
\label{eeq:2bis}
\Thetab_i(j,k) = 
\begin{cases}
1, \; \text{ if } j-1 \leq (k-1) \frac{N_i-1}{\Tau-1} < j
\\
0, \; \text{ otherwise }
\end{cases}
,
\end{equation}
with 
$j = 1, \ldots, N_i-1$, 
$k =  1, \ldots, \Tau $ and $\Tau \geq \max_{i} \left\lbrace N_i \right\rbrace$. With this choice, the corresponding Gramian matrices $\Gb_i$,  are diagonal matrices (Fig.~\ref{fig:transf-matrix}), 
\begin{equation}
\label{eeq:2tris}
    \Gb_i 
    = 
    \Thetab^{T}_{i} \Thetab_{i}
    =
    \diag{\gb_i}
    \;
    \mbox{and}
    \;        
    \diag{
    \Thetab_{i}
    \Thetab^{T}_{i} 
    }
    =
    \oneb_{\Tau}     
    ,
\end{equation}
with $\gb_i \in \mathbb{R}^{N_i}$ and $\oneb_{\Tau} \in
\mathbb{R}^{\Tau}$. Therefore, $\Gb_i$ is invertible and 
the back transformation from the phase to the time domain is given by $\Psib_i = \Gb_i^{-1} \Thetab^{T}_{i}$.

% \begin{equation}
% \label{eeq:3}
% \xib_i = \sigmab_i + \etab_i
% ,
% \end{equation}
From \eqref{eeq:1} and \eqref{eeq:2}, the ECG beats satisfy $\xib_i = \sigmab_i + \etab_i$. As shown in Fig.~\ref{fig:time-and-beats}, in the phase domain the beats have been normalized in lengths and the R-peaks are aligned. Therefore, the phase-domain sample variations are only due to the stochastic inter-beat variations of the ECG beats and noise. As our working model, we assume that the phase domain beats $\xib_i$ to be ensembles of an underlying $\Gc\Pc$
\begin{equation}
    \label{eeq:3bis}
    \xib_i 
    \sim 
    \Nc 
    \left( 
    \mub_{\xi}, \Kb_{\xi} 
    \right)
    .
\end{equation}
Moreover, from the time domain noise assumption and \eqref{eeq:2}, the phase domain noise beats also have a zero-mean normal distribution $\etab_i \sim \Nc \left( \zerob, v_n \Thetab_i \Thetab_i^{T} \right)$. Therefore, the phase domain ECG beats follow $\sigmab_i \sim \Nc \left( \mub_{\xi}, \Kb_{\xi} - v_n \Thetab_i \Thetab_i^{T} \right)$, 
where the model parameters $\mub_{\xi}$ and $\Kb_{\xi}$ can be estimated by the sample mean $\bar{\mub}_{\xi} := B^{-1}\sum_{i=1}^B \xib_i$ and the sample covariance $\bar{\Kb}_{\xi}:= B^{-1}\sum_{i=1}^B (\xib_i - \bar{\mub}_{\xi})(\xib_i - \bar{\mub}_{\xi})^T$, where $B$ is the number of beats. Therefore, the time domain (clean) ECG beats follow a Normal distribution $\ssb_i \sim \Nc \left( \mub_{s_i}, \Kb_{s_i} \right)$
with parameters  
\begin{equation}
\label{eeq:5}
    \mub_{s_i}
    =
    \Psib_i
    \bar{\mub}_{\xi}    
    ,
    \;
    \;
    \Kb_{s_i}
    =
    \Psib_i
    \left(
    \bar{\Kb}_{\xi}
    -
    \vh_n 
    \Thetab_i
    \Thetab_i^{T}
    \right)
    \Psib_i^{T}
    ,    
\end{equation}
where $\vh_n$ represents the noise variance estimate and the covariance matrix corresponding to time domains beats $\xb_i$ is given by
\begin{equation}
\label{eeq:6}
    \Kb_{x_i}
    =
    \Psib_i
    \bar{\Kb}_{\xi}
    \Psib_i
    .
\end{equation}
Finally, the filtered beats are defined as the time domain posterior mean, using \eqref{eeq:5} and \eqref{eeq:6}:
\begin{equation}
\label{eeq:7}
    \hat{\ssb}_i
    =
    \mub_{s_i}
    +
    \Kb_{s_i}
    \Kb_{x_i}^{-1}
    \left(
    \xb_i
    -
    \mub_{s_i}
    \right)
    ,
\end{equation}
In the sequel, we refer to $\mub_{s_i}$ and $\hat{\ssb}_i$ as prior-based and posterior-based $\Gc\Pc$ filter results.
%with the (prior) mean $\mub_{s_i}$ and both covariance matrices $\Kb_{s_i}$ and $\Kb_{x_i}$ set via the phase domain measurements sample mean and covariance respectively. %This approach is fully data-driven and the computations are beat-wise, i.e. involving the inversion and multiplications of small matrices with sizes given by time domain beats lengths in samples.
\subsection{The \texorpdfstring{$\Gc\Pc$}{GP} filter with diagonal covariance matrix}
\label{subsec:the-algorithm}
The direct implementation of the filter in \eqref{eeq:7} requires the inversion of covariance matrices that typically have huge condition numbers. The matrix inversion can be avoided if we consider the diagonal case of $\bar{\Kb_{\xi}}$:
\begin{equation}
\label{eeq:8}
    \bar{\kb}_{\xi} 
    = 
    \diag{\bar{\Kb_{\xi}}}    
    ,
    \;
    \kb_{\eta_i} 
    \stackrel{\eqref{eeq:2tris}}{=}
    \vh_n 
    \oneb_{\Tau}
\end{equation}
%with $\bar{\kb}_{\xi} \in  \mathbb{R}^{\Tau}$ leads to the  Algorithm\;\ref{algo:GPDiagAlgo},
In this case, the corresponding time domain matrices are also diagonal and can be computed via
\begin{equation}
\label{eeq:9}
    \kb_{x_i}
    =
    \left(
    \Thetab_{i}^{T}
    \bar{\kb}_{\xi}
    \right)
    \oslash
    \gb_i^{2}
    ,
    \;
    \;
    \kb_{s_i}
    =
    \left[
    \Thetab_{i}^{T}
    \left(
    \bar{\kb}_{\xi}
    -
    \kb_{\eta_i}
    \right)
    \right]
    \oslash
    \gb_i^{2}
\end{equation}
with $\circ$ and $\oslash$ denoting the Hadamard product and division, respectively (element-wise product and division), $\gb_i^{2} := \gb_i \circ \gb_i$, the time domain (prior) mean computed via
\begin{equation}
\label{eeq:10}
    \mub_{s_i}
    =
    \left(
    \Thetab_{i}^{T}
    \bar{\mub}_{\xi}
    \right)
    \oslash
    \gb_i
    ,
\end{equation}
and the corresponding filter given by
\begin{equation}
\label{eeq:11}
    \hat{\ssb}_i
    =
    \mub_{s_i}
    +
    \kb_{s_i}
    \oslash
    \kb_{x_i}
    \circ
    \left(
    \xb_i
    -
    \mub_{s_i}
    \right)
    .
\end{equation}
The overall algorithm for $\Gc\Pc$ ECG filtering is summarized in Algorithm~\ref{algo:GPDiagAlgo} and is available online in our Git repository \cite{DumitruGitCode}.
\begin{algorithm}[b]
\caption{$\Gc\Pc$ ECG filtering}
\label{algo:GPDiagAlgo}
\begin{algorithmic}[1]
\State $\left\lbrace t_{i{R_i}}\right\rbrace_{i}$ = RPeakDetector($\xb$) 
\;\;[Section~\ref{subsec:R-peak-detection}]
\State $\vh_n$ = NoiseVartianceEstimator($\xb$)\;\;\;[Section~\ref{subsec:model's-hyper}]
\Require{$\xb$, $\left\lbrace t_{i{R_i}}\right\rbrace_{i}$} 
\Ensure{$\left\lbrace \sbh_i \right\rbrace_{i}$}
\Function{GPDiag}{$\xb$, $\left\lbrace t_{i{R_i}} \right\rbrace_{i}$, $\vh_n$} \Comment{$\Gc\Pc$ diagonal filter}
\For {all beats}
\Comment{\textit{phase domain} computations}
    \State compute transformation matrices $\Thetab_i$ via \eqref{eeq:2bis}
    \State compute the vectors $\gb_i = \diag{\Thetab_i^{T}\Thetab_i}$ 
    \State compute $\kb_{\eta_i}$ via  \eqref{eeq:8} 
    \State compute the phase beats $\xib_i$ via \eqref{eeq:2}
\EndFor    
    \State compute \textit{phase domain} sample mean $\bar{\mub_{\xi}}$ 
    \State compute \textit{phase domain} sample variance vector $\bar{\kb}_{\xi}$
\For {all beats} 
\Comment{\textit{time domain} computations}
    \State compute ECG prior mean $\mub_{s_i}$ via \eqref{eeq:10} 
    \State compute  ECG variance $\kb_{s_i}$ via \eqref{eeq:9}
    \State compute  measurements variance $\kb_{x_i}$ via \eqref{eeq:9}
    \State compute the filtered ECG $\hat{\ssb}_i$ via \eqref{eeq:11}
\EndFor        
\EndFunction
\end{algorithmic}
\end{algorithm}

\subsection{Computational cost and model selection}
The direct implementation of a $\Gc\Pc$ filter (without the hereby proposed phase-domain model) would be as follows \cite{Rivet2012,Niknazar2012}:%%%denoisers assuming the ECG as an ensemble of an underlying $\Gc\Pc$ with no additional beats assumptions lead, under the additive model in eq.\,\eqref{eeq:1}, to filters defined as the posterior mean given by 
\begin{equation}
    \label{eeq:100}
    \sbh    
    =
    \mub_{\ssb}
    +
    \Kb_{\ssb}\Kb_{\xb}^{-1}
    \left(
    \xb - \mub_{\ssb}
    \right)    
    ,
\end{equation}
with the computational complexity $\Oc(N^3)$, dominated by the inversion of the measurement covariance matrix $\Kb_{x}$. In this approach the model's hyperparameters are the mean $\mub_{\ssb}$, the covariance matrix $\Kb_{\ssb}$) and the noise variance $v_n$ (or more generally the noise covariance matrix) and optimizing them via classical methods (e.g. maximum evidence, leave-one-out cross validation, \cite[Ch.\ 5]{Rasmussen2005}) adds to the computational complexity. For long ECGs, the application of this model is not possible. Previous research considered the $\Gc\Pc$ beat-wise formulation and adopted a model-based approach to confine the structure of the covariance matrices \cite{Rivet2012,Niknazar2012}, but the choice of the particular model-based mean and kernel function families remains ad-hoc and difficult to justify.

The proposed model infers the $\Gc\Pc$ mean and covariance matrix in a data-driven way, based on the sample mean and covariance matrix from the phase domain \eqref{eeq:5} and \eqref{eeq:6}, and in the diagonal case, Algorithm \ref{algo:GPDiagAlgo}, does not require any inversion.
The fundamental assumption allowing the data-driven computation is the assumption that the phase domain beats $\xib_i$ are ensembles from the same underlying $\Gc\Pc$, \eqref{eeq:3bis}. 

% \subsection{Baseline wander removal}
% \label{subsec:BW-removal}
% The BW is removed via two successively low pass (LP) filters with the cut-off value $k = 0.7$ and the cut-off frequencies set at $f_c = 5.0$ and $f_c = 80.0$ respectively as proposed in \cite{Sameni2007}.

% \subsection{R-peak detection and heartbeat segmentation}
% \label{subsec:R-peak-detection}
% The beats are defined relative to the R-peaks, segmenting the measurements at the midpoints between successive R-peaks, hence the proposed model requires the knowledge of the R-peak positions in the ECG. The R-peak estimation is done using a modified version of the Pan–Tompkins algorithm \cite{PanTompkins1985} for QRS complexes detection. Specifically, the version used in this paper estimates the R-peaks by successively applying: a band pass filter, a saturation filter via the hyperbolic tangent function, a square root moving average filter and a thresholding filter with the parameter. This particular R-peak detector was first proposed in \cite{Sameni2007}.

\subsection{Hyperparameter selection}
\label{subsec:model's-hyper}
The number of phase domain beat samples $\Tau$ is chosen greater than the longest beat in the time domain; this allows the choice of the transformation and back transformation matrices such that the \textit{time-phase-time} transition can be done without (transformation) errors.
%%%The mean and kernel functions are inferred via the sample mean and covariance, computed in the phase domain and projected back in the time domain. The model's hyperparameter is 
The noise variance $\vh_n$ can be computed via maximum evidence or practically from the baseline segment of the ECG beats, where the heart is electrically silent and only the noise is exhibited in the ECG.

\iffalse
\subsection{Noise variance estimaton}
\label{subsec:noise-variance-estimation}
Because the additive noise is
assumed to be spectrally white, a straightforward way to estimate it is via the prior mean, \eqref{eeq:10} (which depends on the \textit{phase domain} sample mean $\bar{\mub}_{\xi}$) and inferring the noise variance from the difference $\xb - \mub_{s}$. Figure \ref{fig:noise-var-estimation} shows the mean ratio between the real standard deviation used in simulations and the estimated one over the PhysioNet QT Database computed for SNR between -5dB and 30dB with a 5db step.
\begin{figure}
\centering
\includegraphics[width=0.95\columnwidth]{Figs/DiagQT_UnknVar_Beta300VarEst}
\caption{The mean ratio between the real standard deviation used in simulations and the estimated one over the PhysioNet QT Database computed for SNR between -5dB and 30dB with a 5db step.}
\label{fig:noise-var-estimation}
\end{figure}
\fi

% Section 3 : Results
\section{Results}
\label{sec:results}

\subsection{Baseline wander removal}
\label{subsec:BW-removal}
The BW is removed via two successively zero-phase first order forward-backward lowpass filters (\texttt{filtfilt} in MATLAB/Python SciPy) with cut-off frequencies set at $f_c=$ 5.0\,Hz and $f_c= $ 80.0\,Hz, respectively. While the resulting passband frequency range is rather narrow and eliminates some ECG-related components, it enables us to assess the filtering performance for the dominant ECG frequency band.

\subsection{R-peak detection and heartbeat segmentation}
\label{subsec:R-peak-detection}
The proposed filter requires the ECG R-peaks. The beats are defined relative to the R-peaks, segmenting the measurements at the midpoints between successive R-peaks. The R-peak estimation is done using a modified version of the Pan–Tompkins algorithm \cite{PanTompkins1985}. Specifically, the version used in this paper estimates the R-peaks by successively applying a band pass filter, an outlier saturation filter via the hyperbolic tangent function, a square root moving average filter and a thresholding.%\cite{Sameni2007}.

\subsection{Evaluation}
\label{subsec:evaluation}
The PhysioNet QT Database (QTDB) \cite{Laguna1997} is used to evaluate the developed filter.
%The data used for this section is the \textsc{QTDataBase} file from \textsc{PhysioNet} database \cite{SameniOSET314}.
QTDB consists of 15 minutes 2-lead ECGs sampled at $f_s =$ 250\,Hz. The baseline wander was removed as detailed in Section~\ref{subsec:R-peak-detection}. The required software for preprocessing and R-peak detection were adopted from the Open-Source Electrophysiological Toolbox (OSET) \cite{SameniOSET314}. 

% \subsubsection{Benchmark}
% \label{subsubsec:benchmark}
The benchmark filter is a wavelet denoiser with a \mbox{\textit{Symlet--5}} mother wavelet, soft thresholding, Stein's unbiased risk estimate (SURE) shrinkage rule, rescaling using a single-level noise level estimation and four levels of decomposition. In a previous study, this combination was proved to outperform other ECG filtering schemes \cite{Sameni2017}. The filter evaluation is measured in terms of SNR improvement and QT-interval estimation error. %, which was empirically found to be sufficient to represent the average beats.
%The corner case version of the proposed data-driven $\Gc\Pc$-based filter is considered via both the $\Gc\Pc$  posterior‐based filter  \eqref{eq:filter-phase-cc} and the $\Gc\Pc$ prior‐based filter, which just corresponds to $\sbh = \Theta^{T}\Theta$ and serves as benchmark for very noisy inputs. 

\subsection{SNR improvement performance}
\label{subsec:subsec52}
\begin{figure*}
    \centering
    \subfigure[input SNR = 0\,dB]{\includegraphics[trim=0in 0in 0in 0in, clip, width=0.3\linewidth]{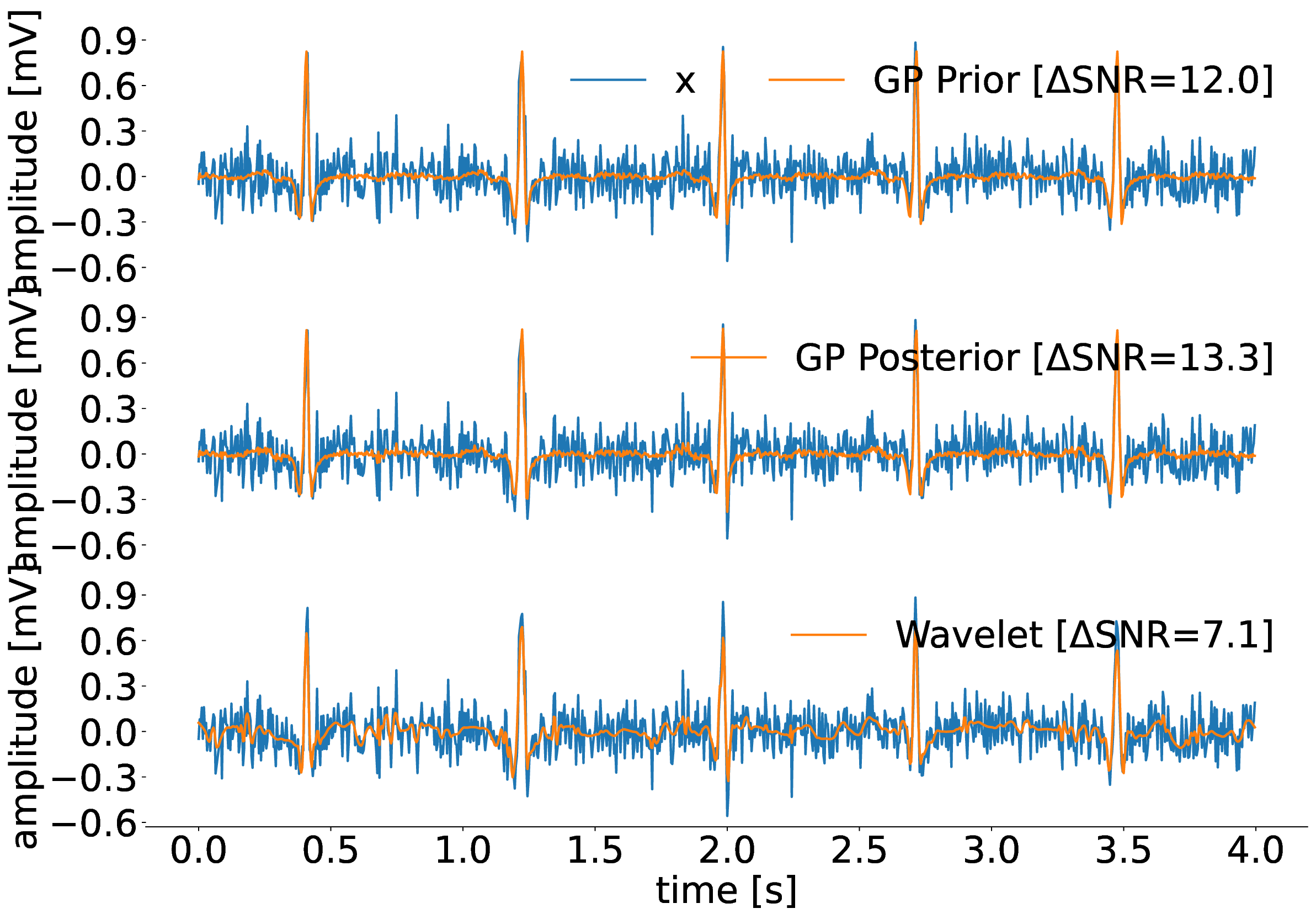}
    \label{fig:0dB}}
    \subfigure[input SNR = 5\,dB]{\includegraphics[trim=0in 0in 0in 0in, clip, width=0.3\linewidth]{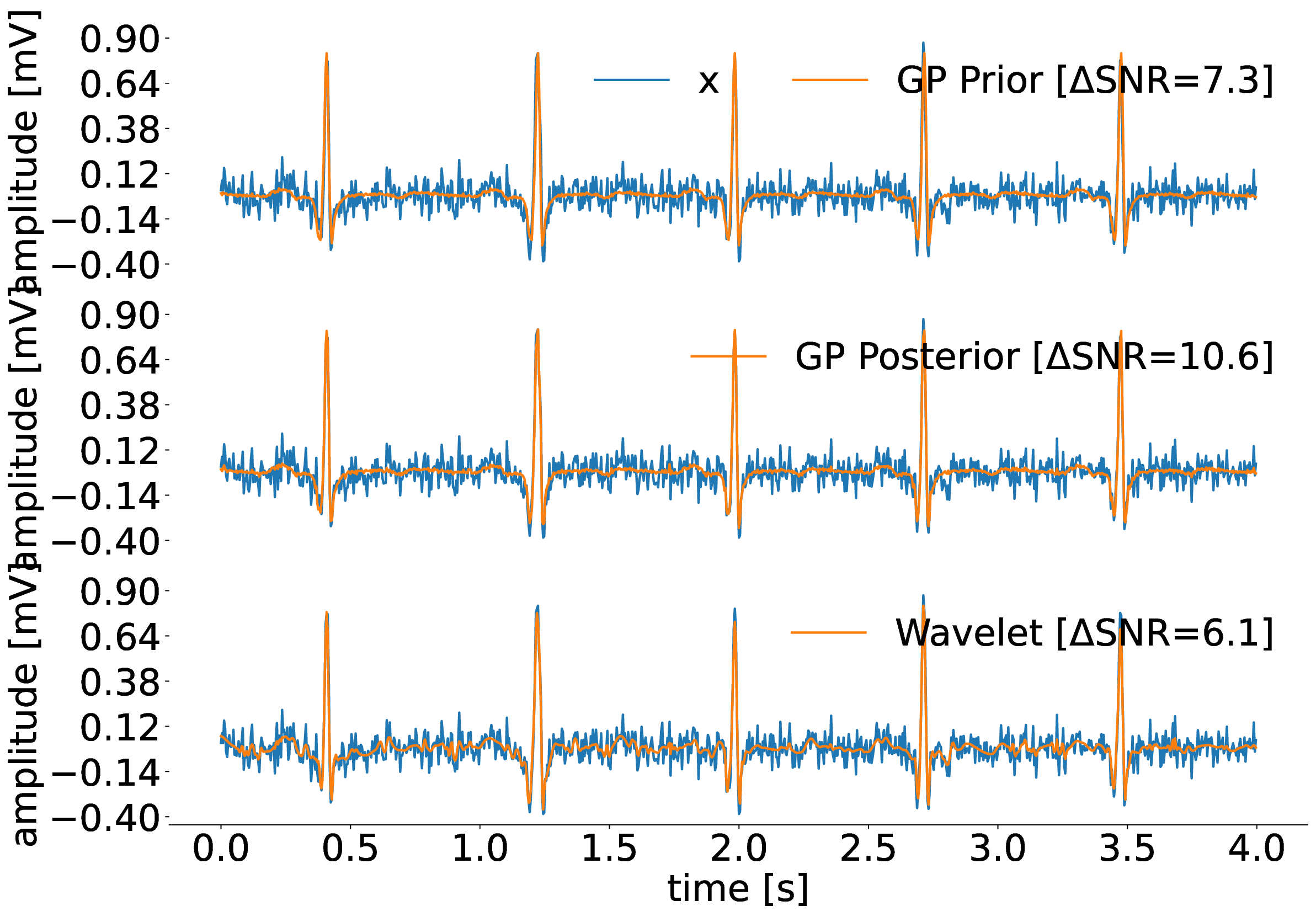}
    \label{fig:5dB}}    
    \subfigure[input SNR = 10\,dB]{\includegraphics[trim=0in 0in 0in 0in, clip, width=0.3\linewidth]{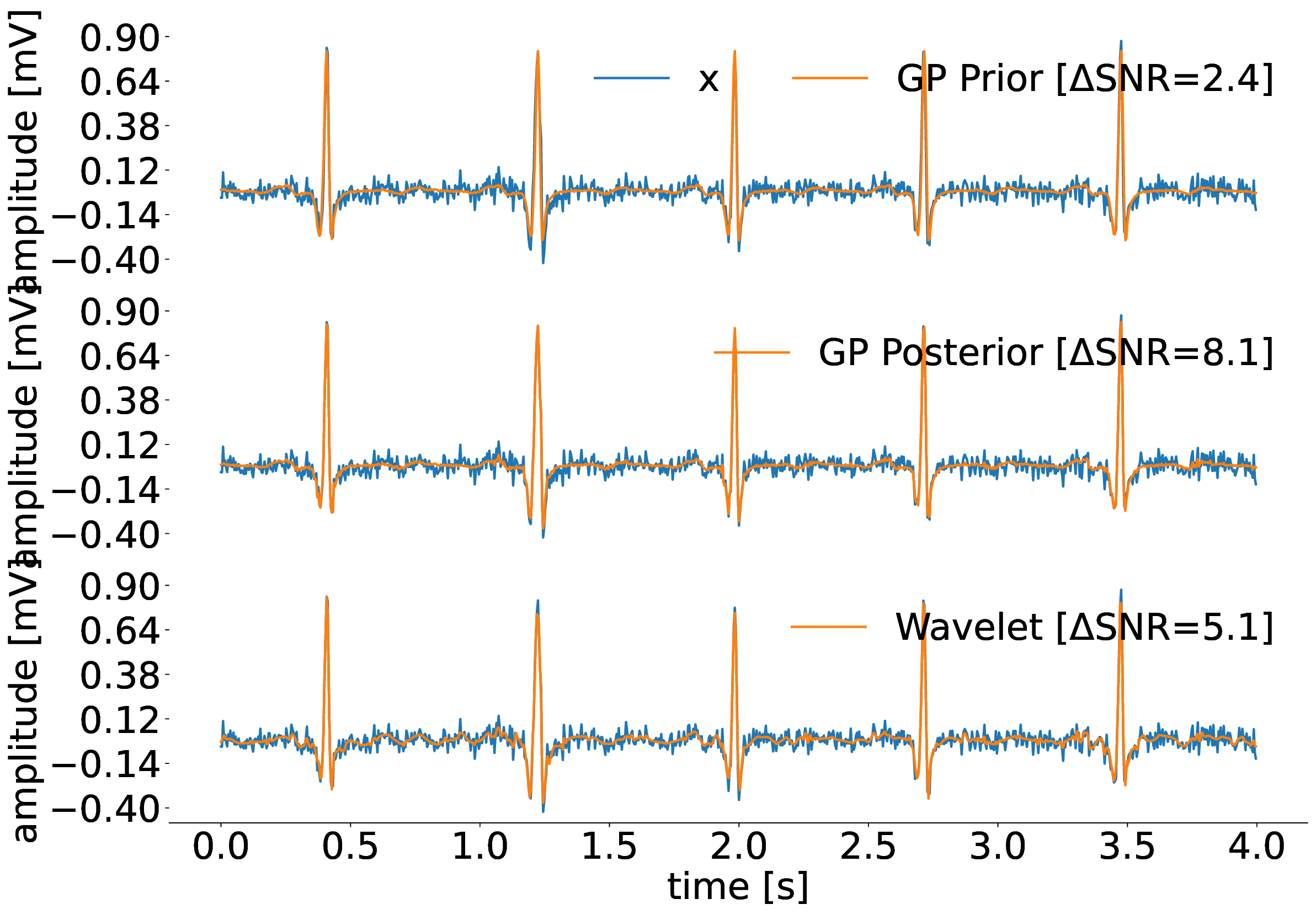}
    \label{fig:30dB}}
\caption{The \texttt{sel100} recording from the PhysioNet QTDB \cite{Laguna1997}. From top to bottom the measurements $\xb$ vs. the prior estimate \eqref{eeq:10}, the posterior estimate \eqref{eeq:11}, and the wavelet denoiser (Section~ \ref{subsec:evaluation}), at different input SNR levels. The post-filtering SNR improvement is noted in each case.}
\label{fig:filter_results}
\end{figure*}

\begin{figure}
\centering
\includegraphics[width=0.98\columnwidth]{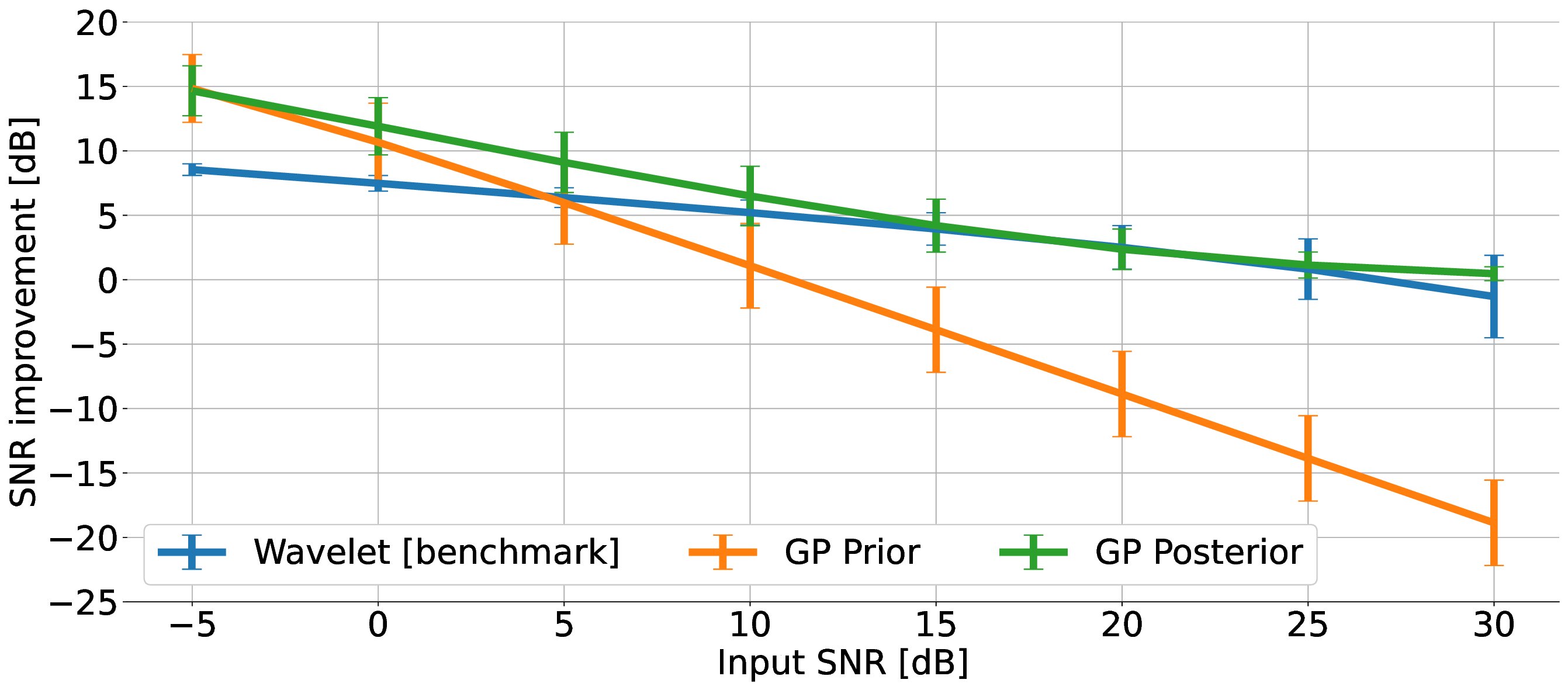}
\caption{Mean and standard deviation SNR improvement using the proposed $\Gc\Pc$ filter and the benchmark wavelet denoiser \cite{Sameni2017} across all samples of the PhysioNet QTDB \cite{Laguna1997}, in leads I and II, with 5 repetitions using different noise instances per record.}
\label{fig:results-SNR-curves}
\end{figure}

%%%For performance assessment,
The ECG records were contaminated by additive white Gaussian noise at SNR levels ranging from --5 to 30\,dB, in 5\,dB steps. An example of the noisy and filtered ECG are shown in Fig.~\ref{fig:filter_results}. The average and standard deviation of the SNR improvement is reported for each noise level, for the proposed and benchmark methods in Fig.~\ref{fig:results-SNR-curves}. %and the top part of Table\,\ref{tab:SNR-improvement-and-QT-means}. 
Accordingly, the proposed posterior-based filter improves the SNR for every level of noise tested and outperforms the prior-based and the benchmark filter for all tested levels of noise. %Note that the implemented $\Gc\Pc$ filters correspond to the corner case \eqref{eq:filter-phase-cc}, with no inversion required.

\subsection{Clinical parameters preservation}
\begin{figure}
\centering
\includegraphics[width=0.98\columnwidth]{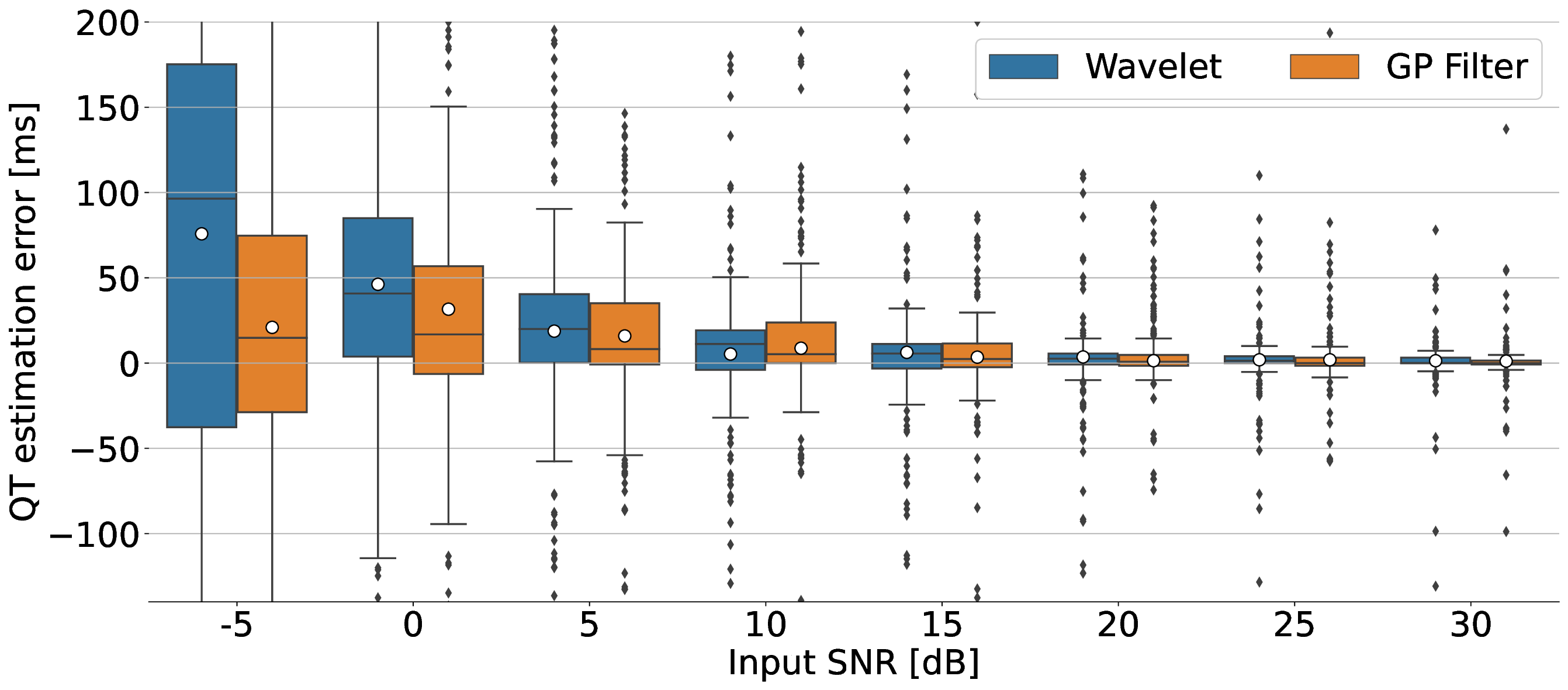}
\caption{The median and the interquartile range for $\Delta$QT estimations corresponding to the proposed and benchmark filters across all samples of the PhysioNet QTDB \cite{Laguna1997}. %The $\Gc\Pc$ filtered ECGs have smallerestimation error bias and variance compared with the wavelet denoiser.
}
\label{fig:results-QT-estim-curves}
\end{figure}

%As proof of concept,
The accuracy of QT-interval estimation is considered to test the quality of the proposed methods for clinical ECG parameters. For this, the QT-interval estimation error ($\Delta$QT) between the QT-interval estimated from the filtered ECG and the QT-interval estimated from the noiseless ECG is measured and compared between the benchmark and the proposed method at variable input noise levels. The QT-interval estimation method used is adopted from \cite{Li2022}. 
%\textcolor{red}{[EDITED UP TO HERE - Reza]}
Fig.~\ref{fig:results-QT-estim-curves} shows the median and the interquartile range (IQR) of $\Delta$QT for the benchmark wavelet and the proposed filter, measured over QTDB. Accordingly, compared with the benchmark method, the $\Gc\Pc$ posterior filter is reducing the median error for all levels of input noise. %The median and IQR of $\Delta$QT is reported on the bottom of Table~\ref{tab:SNR-improvement-and-QT-means}.
%%%The mean $\Delta$QT corresponding to the $\Gc\Pc$-posterior filter is under 5\,ms for every level of noise tested. 

% Section 4 : Conclusion
\section{Discussion and conclusion}
\label{sec:conclusion}
In this work we addressed the problem of ECG denoising via a data-driven based $\Gc\Pc$ model, with beat-wise computations. Compared with the existing non-parametric ECG filters, the proposed filter makes no \textit{ad hoc} assumptions about the $\Gc\Pc$ model and can be used for ECG records of arbitrary length, since the computational cost has been significantly reduced as compared with conventional $\Gc\Pc$ filters. The proposed filter is efficient in terms of SNR improvement, outperforming the benchmark performances for all tested noise levels (Fig.~\ref{fig:results-SNR-curves}) and also clinically, with an improved QT-interval estimation error compared with the benchmark wavelet denoiser, for all tested levels of noise (Fig.\,\ref{fig:results-QT-estim-curves}). Another advantage of the proposed filter is its Bayesian formulation, which allows us to quantify the filter's uncertainty (via the estimated variances). It also provides a framework that allows for synthetic ECG generation via data-driven learned parameters, which can be used in generative models for producing synthetic ECG records for data greedy machine learning and deep learning applications.

In future studies, the fundamental assumption of the model, namely the same underlying Gaussian distribution for all the beats in the phase domain can be relaxed, by clustering the beats and assuming different underlying distributions for the beats in each cluster. Also, comparison with expert annotated QT-interval (and other clinical parameters) is required and statistical hypothesis testing should be performed to investigate if the differences are statistically insignificant. The proposed filter requires the R-peaks for aligning the ECG beats in the phase-domain, which requires investigating to what extend the filtering performance is susceptible to mis-detection of the R-peaks and morphological variations due to ectopic beats.
The Python codes corresponding to the Algorithm~\ref{algo:GPDiagAlgo} and the reported results are available in \cite{DumitruGitCode}.

%\subsection{Discussion}
%\label{subsec:discussion}
%The fundamental idea in the proposed model was to consider the representation of the ECG in the phase domain. %%%The choice for the transformation matrix $\thetab$ simplifies the transformation choices in \eqref{eq:choices} and illustrates well the data-driven approach but it is limiting the values of $\Tau$ to be at most equal to shortest beat number of samples. The model can be easily generalized - overcoming this limitation - and investigating the influence of the parameter $\Tau$ and the gain in terms of SNR improvement and clinical parameters error estimation induced by using the full sample covariance matrix (\ie \eqref{eq:filter-phase} instead of \eqref{eq:filter-phase-cc} is a short term future work. In addition to this, the reduced computational cost of the proposed method has to be assessed by comparing it to the one corresponding to the state-of-the-art methods.

% Section 5 : Disclosures
\section{Acknowledgements}
\label{sec:acknowledgements}
The authors acknowledge support from the National Institute of Biomedical Imaging and Bioengineering under the NIH grant R01EB030362, and the National Center for Advancing Translational Sciences under the NIH Award UL1TR002378.

% BIBLIOGRAPHY
% \Urlmuskip=0mu plus 1mu\relax
\bibliographystyle{IEEEtran}
\bibliography{biblio}
\end{document}